\documentclass[prl,twocolumn,showpacs,preprintnumbers,amsmath,amssymb]{revtex4}

\usepackage{graphicx}

\begin{document}

\title{Complex evolution of the electronic structure from
polycrystalline to monocrystalline graphene: generation of a new Dirac
point}

\author{Joice da Silva Ara\'ujo} 

\affiliation{Departamento de F\'{\i}sica, Universidade Federal de
Minas Gerais, CP 702, 30123-970, Belo Horizonte, MG, Brazil}

\author{R. W. Nunes}
\email{rwnunes@fisica.ufmg.br}
\affiliation{Departamento de F\'{\i}sica,  Universidade Federal de Minas
Gerais, CP 702, 30123-970, Belo Horizonte, MG, Brazil}

\date{\today}


\begin{abstract}
First calculations, employed to address the properties of
polycrystalline graphene, indicate that the electronic structure of
tilt grain boundaries in this system display a rather complex
evolution towards graphene bulk, as the tilt angle decreases, with the
generation of a new Dirac point at the Fermi level, and an
anisotropic Dirac cone of low energy excitations. Moreover, the usual
Dirac point at the {\bf K} point falls below the Fermi level, and
rises towards it as the tilt angle decreases. Further, our
calculations indicate that the grain-boundary formation energy behaves
non-monotonically with the tilt angle, due to a change in the the
spatial distribution and relative contributions of the bond-stretching
and bond-bending deformations associated with the formation of the
defect.
\end{abstract}

\maketitle

Graphene - an isolated layer of $sp^{2}$-bonded carbon atoms arranged
in a honeycomb structure - was until very recently, a ``theoretical''
reference system for the study of the properties or the ``real''
$sp^2$-bonded carbon forms, such as fullerenes, nanotubes, and
graphite~\cite{rmp,novos07,nakada,wallace}. Since the recent report of
the isolation of a stable single-atom-thick carbon layer, by
exfoliation of graphite~\cite{novos04}, graphene itself has occupied
the center stage of Materials Physics, as a paradigmatic system for
``relativistic'' condensed-matter phenomena, as well as a promising
material for Nanoelectronics, due to its exceptional electronic
properties. Graphene is a null-gap semiconductor with a vanishing
density of states at the Fermi level, and electronic bands that are
linear and isotropic within $\sim$1 eV from the Fermi level. This
linearity and the presence of two sub-lattices imply that charge
carriers in graphene effectively behave as massless chiral
``relativistic'' particles, being described by Dirac's
equation.~\cite{rmp,novos07,wallace} Due to the chiral nature of the
electronic excitations, which leads to the absence of backscattering,
graphene holds ballistic charge transport on the microscale, even at
room temperature, and with high concentrations of defects and
impurities.~\cite{rmp,novos07,novos04,kats06} In the last few years,
scientific interest in graphene has rapidly intensified, and the
material is expected to play a major role in Nanoelectronics in the
future.

Presently, common synthesis routes for graphene are the original
exfoliation method,~\cite{novos04} that produces monocrystalline
graphene samples, and epitaxy, mostly on SiC~\cite{deheer} and
metallic substrates~\cite{sinitsyna}. Recent works on
epitaxially-grown graphene report the occurrence of superstructures
interpreted as Moir\'e patterns~\cite{laura,datta} based on STM, AFM,
and STS measurements. Moir\'e patterns and superstructures, associated
with layer stacking, as well as the occurrence of bulk and surface
grain boundaries, are topics of prominence in the physics of
highly-oriented pyrolytic graphite (HOPG)
itself.~\cite{simonis,gan,pong,varchon,cervenka} Grain boundaries (GB)
are among the most commonly occurring extended defects in HOPG,
because of its polycrystalline character~\cite{cervenka}. For large
scale graphene production and application, it is expected that
synthesis methods will be epitaxy based, being quite conceivable that
polycrystalline samples will be produced. Indeed, the occurrence of GBs
on the graphene layer has been recently reported,~\cite{laura}
speculated as a probable source of long-range electronic perturbations
in graphene on SiO$_{\rm 2}$,~\cite{datta} and further, the lower
carrier mobility of epitaxial graphene (when compared with exfoliated
samples), in a macroscopic-size graphene field-effect transistor, has
been tentatively assigned to the electronic perturbations associated
with GBs.~\cite{gong}
\begin{figure}
\includegraphics[width=8.5cm]{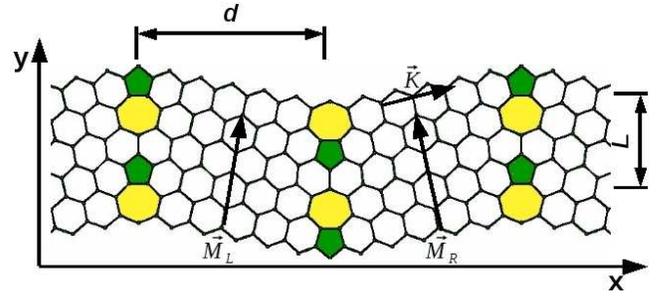}\\
\caption{(Color online) Grain boundary geometry: the relative
orientation between grains is defined by the angle
between vectors $\vec{M_L}$ and $\vec{M_R}$. The distance between
the two grain boundaries in the supercell is $d$, and $L$ is the
period of the pentagon-heptagon pattern along the grain boundary.}
\label{cell}
\end{figure}

Previous works have addressed the electronic properties of disordered
graphene, but these have focused on the effects of point defects,
corrugation, and extended edge states in graphene ribbons, drawing
only speculative conclusions in what regards the electronic states of
GBs.~\cite{rmp} In this scenario, addressing properties of GBs in
graphene is of primary relevance. In the present work, we employ first
principles calculations to examine the energetics and electronic
properties of GBs in graphene. We focus on the structural GB model
proposed by Simonis {\it et al.},~\cite{simonis} who observed a
large-angle tilt boundary on the surface of HOPG, on STM
experiments. These authors proposed that, in the absence of stress,
the observed GB consists of a periodic structure that can be described
as a regular succession of pentagon-heptagon pairs, as shown in
Fig.~\ref{cell}. Based on this model, we investigate three GBs of
different periodicities along the boundary, hence with different
relative orientation between the grains.

We find that, while large-angle tilt GBs do not introduce localized
states at the Fermi level in graphene, various resonance peaks appear
in the density of states of the material, over a broad energy range,
starting at energies of $\sim$0.3 eV from the Fermi level, in
agreement with recent experimental work.~\cite{cervenka} More
importantly, the changes in electronic structure with the GB tilt
angle indicate a non-trivial evolution towards graphene bulk, as the
GB angle decreases: we observe the generation of a new Dirac point at
the Fermi level, which lies on a line that evolves towards the {\bf
$\Gamma$-M} direction of the graphene Brillouin zone (BZ), with the
usual Dirac point at the {\bf K} point falling below the Fermi level,
and rising towards it as the tilt angle decreases. The Dirac cone of
low-energy excitations around this new Dirac point is non-isotropic,
with the effective ``speed-of-light'' depending on the direction in
the BZ away from the Dirac point. Furthermore, our calculations
indicate that, within the structural model we consider, the GB
formation energy does not behave monotonically with the period of the
GB (or, equivalently, with the tilt angle). A Keating analysis of the
elastic energy, associated with the formation of the defect, indicates
that a change in the the spatial distribution and relative
contributions of the stretching and bending deformations leads to the
non-monotonicity indicated by our results.
\begin{table}
\caption{Geometric parameters and formation energy per unit length
$E_f$ (in eV/\AA), of GB supercells containing $N_a$ atoms. $\alpha$
is the tilt angle and $L$ is the GB period (in \AA), as indicated in
Fig.~\ref{cell}. $E_{el}^{str}$ and $E_{el}^{bend}$ are the
contributions to the Keating-model total elastic energy per unit
length $E^{el}$}.
\begin{tabular}{cccccccc}
\hline
& \;\;$N_a$\;\; & \;\;$\alpha$\;\; & \;\;$L$\;\; & \;\;$E_f$\;\; 
& \;\; $E_{el}$\;\; & \;\;$E_{el}^{str}\;\;$ & \;\;$E_{el}^{bend}\;\;$\\
\hline
GB1 &  72 & $21.8^{0}$  & 6.6 & 0.33 &0.42 &0.10 &0.32\\
GB2 & 120 & $13.3^{0}$ & 10.9 & 0.42 &0.47 &0.15 &0.31\\
GB3 & 168 &  $9.6^{0}$ & 15.2 & 0.40 &0.41 &0.15 &0.27\\
\hline
\end{tabular}
\label{energy}
\end{table}

All calculations are performed using Kohn-Sham density functional
theory~\cite{Kohn}, the generalized-gradient approximation
(GGA)~\cite{Kleinman} for the exchange-correlation term, and
norm-conserving Troullier-Martins pseudopotentials \cite{Troullier},
to describe the electron-ion interaction. We use the LCAO method
implemented in the SIESTA code~\cite{Siesta}, with a double-zeta
pseudo-atomic basis set plus polarization orbitals, with an energy
cutoff of 0.01 Ry. Structural optimization is performed until the total
force on each atom is less than 0.02 eV/\AA. In order to simulate an
isolated honeycomb sheet, we use supercells that are periodic along
the graphene plane, and are surrounded by a 33 \AA\ vacuum region,
such that the interactions between each layer and its periodic images
are negligible.

Periodicity along the graphene plane requires the supercell to contain
two GBs of opposite tilt angles (a GB and the corresponding
``anti-GB''), as shown in Fig.~\ref{cell} for the GB1 geometry. The
experimental value in Ref.~\cite{simonis} for the relative orientation
between grains is $21^\circ$, defined here by the angle $\alpha$
between vectors $\vec{M_L}$ and $\vec{M_R}$, drawn respectively on
the left and right grains adjacent to the GB, as shown in
Fig.~\ref{cell}. This model can be extended to GBs with smaller tilt
angles, by adding lines of hexagons, such that the period $L$ of the
pentagon-heptagon pattern along the GB increases. In the $L
\rightarrow\infty$ ($\alpha\rightarrow 0$) limit, we recover the
perfect single-crystal graphene lattice. We study three different GB
geometries, with the theoretical values for $L$ and $\alpha$ indicated
in Table~\ref{energy}. GB1 is the model proposed in
Ref.~\cite{simonis}, with $\alpha=21.8^\circ$ and $L=6.6$~\AA; GB2 has
$\alpha = 13.3^\circ$ and $L=10.9$~\AA; and for GB3 $\alpha =
9.6^\circ$ and $L=15.2$~\AA. In order to ensure that we simulate the
properties of an isolated GB, we consider supercells with increasing
distances $d$ between the GBs and their periodic images. Formation
energy results are converged for $d=14.9$~\AA. The geometric
parameters $\alpha$, $d$, and $L$ are shown in Fig.\ref{cell}, with
values for $\alpha$ and $L$ given in Table~\ref{energy}.
\begin{figure}
\includegraphics[width=8.0cm]{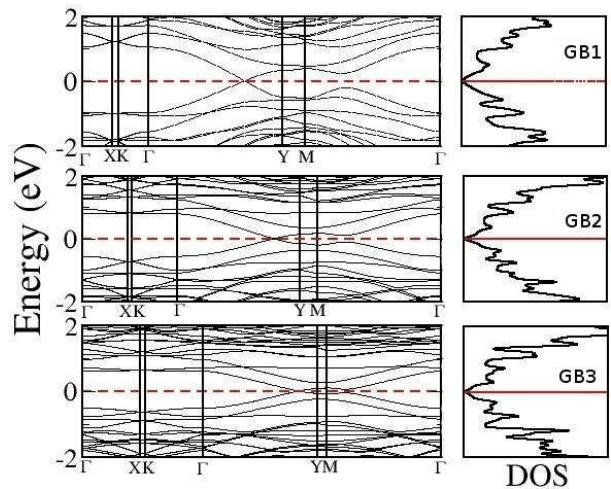}\\
\caption{Band structure and density of states for (a) GB1, (b) GB2,
and (c) GB3 geometries. The Fermi level is indicated by the dashed
line. Brillouin-zone lines are shown in Fig.~\ref{cone}.}
\label{band}
\end{figure}

\begin{figure}
\includegraphics[width=8.0cm]{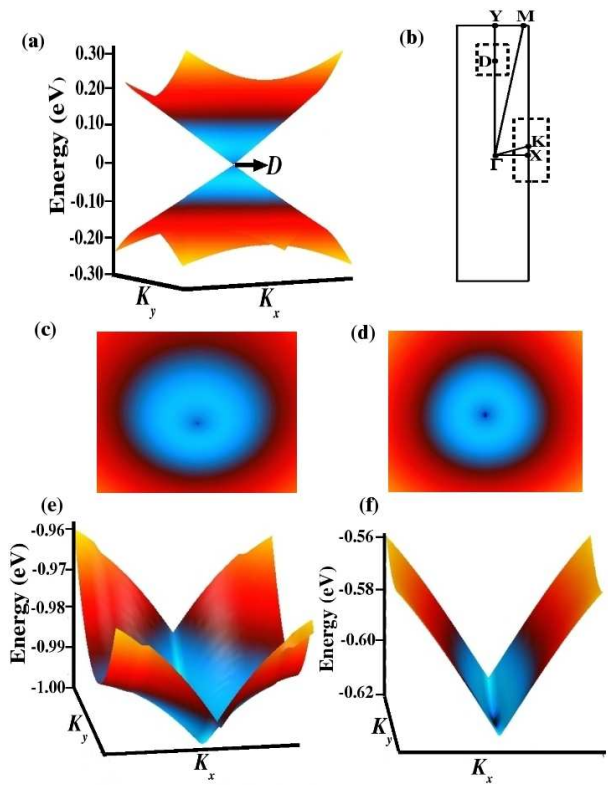}\\
\caption{(Color online) Electronic energy dispersion for
polycrystalline graphene. (a) The anisotropic Dirac cone around the
{\bf D} point, for the GB1. (b) The Brillouin-zone directions and
regions for the GB1. (c) and (d) The color-coded constant-energy
isolines (with the energy increasing from violet to red) for the GB1
and GB3, respectively. (e) and (f) The wedge-shaped dispersion around
the {\bf K-X} line shown in (b).}
\label{cone}
\end{figure}

The band-structure of the GB1 ($\alpha = 21.8^\circ$), is shown in
Fig.~\ref{band}(a). The Brillouin zone for this GB supercell is shown
in Fig.~\ref{cone}(b). The {\bf $\Gamma-$Y} direction is along the GB,
and the {\bf $\Gamma-$X} direction is perpendicular to it. In
Fig.~\ref{cone}(b), we also indicate lines we denoted as {\bf
$\Gamma-$M} and {\bf $\Gamma-$K}, which correspond to these
high-symmetry directions in a single-crystal graphene sheet with the
orientation of the grain on the right side of the GB. {\bf $\Gamma-$M}
(and {\bf $\Gamma-$Y}) and {\bf $\Gamma-$K} (and {\bf $\Gamma-$X})
converge to the corresponding directions in the $\alpha\rightarrow 0$
limit, where a single-crystal graphene sheet is recovered. A mirror
plane, perpendicular to the graphene sheet and through the geometric
center of the GB, relates the {\bf $\Gamma-$M} and {\bf $\Gamma-$K}
lines of the two adjacent grains. Note that the Fermi level or Dirac
point occurs on the {\bf $\Gamma-$Y} line, in the point marked as $D$
in Fig.~\ref{cone}(b), and that band crossings, which lie below the
Fermi level, occur at the {\bf X} and {\bf K} points.

Thus, we observe the generation of a new Dirac cone~\cite{park}, on
the {\bf $\Gamma-$Y} direction, which is anisotropic, with a Fermi
velocity (the effective ``velocity of light'') that depends on the
direction from the Dirac point in ${\bf k}$-space. This Dirac cone is
shown in Fig.~\ref{cone}(a), for the region around the {\bf D} point
indicated in Fig.~\ref{cone}(b)). The energy isolines are shown
Fig.~\ref{cone}(c), where the anisotropy of the cone is clearly
displayed: the more (less) elongated is the isoline along a given
direction, the smaller (larger) is the Fermi velocity along that
direction. The band-structures for the GB2 ($\alpha = 13.3^\circ$) and
GB3 ($\alpha = 9.6^\circ$) are shown in Figs.~\ref{band}(b) and (c),
respectively. Note again the occurrence of a Dirac point at the Fermi
level, on the {\bf $\Gamma-$Y} line, in both cases. Both display
anisotropic Dirac cones as well. The isolines for the GB3, shown in
Fig.~\ref{cone}(d), are less anisotropic than in the GB1 case. In
Table~\ref{veloc} we include the Fermi velocities for the Dirac cones
of the three GBs, along the indicated directions.

The nature of the electronic dispersion in the region around the {\bf
X} and {\bf K} points, also indicated in Fig.~\ref{cone}(b), are shown
in Fig.~\ref{cone}(e) and (f), for the GB1 and GB3, respectively.  For
the GB3, there is very little dispersion along the {\bf X-K}
direction, resulting in a straight wedge cutting through this line,
while in the GB1, the shape of the energy dispersion is also
wedge-like, but a ``flowery'' shape develops at higher
energies. Furthermore, the band crossings at the {\bf X} and {\bf K}
points move up towards the Fermi level, as $\alpha$ decreases, which
is consistent with the $\alpha\rightarrow 0$ limit, where the Fermi
level occurs at the {\bf K} point. The energy difference between the
Dirac point at {\bf D} and the {\bf K} point is 1.00 eV for the GB1,
0.73 eV for the GB2, and 0.64 eV for the GB3.

These results indicate a very complex evolution of the band-structure
of polycrystalline graphene with the GB angle. Since fivefold and
sevenfold topological defects constitute the core of low-energy
dislocations in graphene,~\cite{vozmed} they are probably ubiquitous
in any realistic model of GBs in this material. Hence, the occurrence
of a new Dirac point along the GB direction may prove a robust feature
of polycrystalline graphene. We note that previous works have found
that, while pentagon-heptagon pairs and the related SW defect
introduce no resonant states {\it at the Dirac point} in graphene,
resonance peaks appear in the density of states starting at a few
tenths of an eV from the Fermi level, a result that we have also
reproduced with the {\it ab initio} method employed here. Our GB
electronic structure calculations show, however, that the periodic
superstructure formed by these dislocations along the GB line shares
with perfect graphene the vanishing gap and the Dirac-like nature of
electronic excitations, but in a rather complex structure, with
direction-dependent Fermi velocities, the generation of a new Dirac
point on the GB direction, and a wedge-like dispersion around the {\bf
K} point that should evolve towards the graphene Dirac cone as
$\alpha\rightarrow 0$. We expect charge transport in polycrystalline
graphene to reflect the anisotropic structure of the Dirac cones we
obtain in our calculations.
\begin{table}[t!]
\caption{Fermi velocities (in 10$^6 {\rm m/s}$) for the GB Dirac
electron cones, along the indicated directions.}
\begin{tabular}{ccccccc}
\hline
&$\;\;\;\;-\hat{y}\;\;\;\;$ 
&$\;\;\;\;\;\hat{y}\;\;\;\;$ 
&$\;-\hat{x}-\hat{y}\;\;$ 
&$\;\;\hat{x} +\hat{y}\;\;$
&$\;\;\;\;-\hat{x}\;\;\;\;$  
&$\;\;\;\;\;\hat{x}\;\;\;\;$ \\
\hline
GB1 &0.73 &0.52 &1.06 &0.69 &0.60 &0.60\\
GB2 &0.64 &0.43 &0.91 &0.63 &0.52 &0.52\\
GB3  &0.55 &0.45 &0.53 &0.55 &0.57 &0.57\\
\hline 
\end{tabular}
\label{veloc}
\end{table}

We turn now to the GB energetics. The GB formation energy per unit
length is given by $E_f=\left(E_{GB}-E_{bulk}\right)/2L$, where
$E_{GB}$ and $E_{bulk}$ are the calculated total energies of the GB
and bulk graphene supercells, respectively. The factor of two on the
right-hand side accounts for the presence of two GBs in the cell.
\begin{figure}
\includegraphics[width=9.0cm]{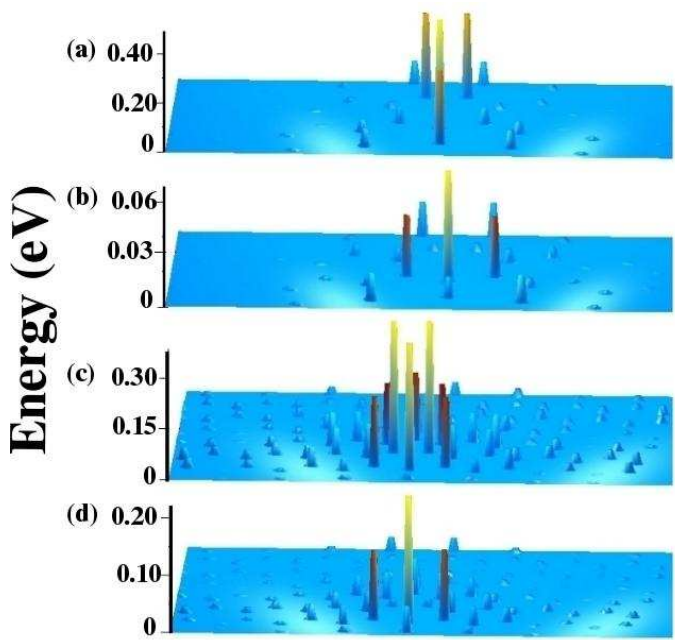}\\
\caption{(Color online) Spatial distribution of bond-bending and
bond-stretching energies of grain boundaries in graphene. (a) Bend for
GB1. (b) Stretch for GB1. (c) Bend for GB3. (d) Stretch for GB3.}
\label{keating}
\end{figure}

Results in Table~\ref{energy} show that $E_{f}$ varies
non-monotonically with the GB period. Among the three GBs, GB1 with a
tilt angle of $21.8^\circ$ has the smallest formation energy, while
GB2 with $\alpha = 13.3^\circ$ has the highest $E_f$ value. We expect
that $E_{f}\rightarrow 0$ as $\alpha\rightarrow 0$. Our results
indicate that $E_{f}$ initially increases, before reaching this
asymptotic limit. Given the absence of broken bonds in the GB
geometries, $E_{f}$ is primarily of elastic nature. In order to
understand this non-monotonic behavior, we use a Keating
model to analyze the bond-bending $E_{el}^{bend}$, and
bond-stretching $E_{el}^{str}$ contributions to the GB elastic energy
$E_{el}$. The results are shown in Table~\ref{energy}. For the GB1 we
observe that $E_{el}^{bend}$ is a factor of 3.2 larger than
$E_{el}^{str}$, while for the GB2 and GB3 structures the
$E_{el}^{bend}/E_{el}^{strch}$ ratio decreases to 2.1 and 1.8
respectively. These results remain essentially unaltered for different
choices of the ratio between the stretching and bending energy
parameters of the Keating model.  Moreover, the spatial distribution
of $E_{el}^{bend}$ and $E_{el}^{strech}$ changes as $\alpha$
decreases, as shown in Fig.~\ref{keating}. For the GB1,
$E_{el}^{bend}$ is largely concentrated on the atoms at the very core
of the boundary, mainly in the pentagon-heptagon pairs, while
$E_{el}^{str}$ is a little more spread out towards the interior of the
grains. For GB3, we see that both $E_{el}^{bend}$ and $E_{el}^{str}$
have significant contributions from atoms in the interior of the
grain, mostly on the hexagon lines that at the pentagon-heptagon at
the boundary. Thus, by concentrating the elastic energy on bending
distortions at the defect core, the GB1 structure relaxes to a lower
energy state than GBs with smaller angles, which leads to the
non-monotonic behavior of $E_{f}$.

To conclude, {\it ab initio} calculations indicate that the electronic
structure of tilt GBs in graphene display a complex evolution towards
graphene bulk, as the GB angle decreases, with the generation of a new
Dirac point at the Fermi level, lying on a line that evolves towards
the {\bf $\Gamma$-M} direction of the graphene Brillouin zone, at the
vertex of an anisotropic electronic-energy cone. Moreover, the usual
Dirac point at the {\bf K} point falls below the Fermi level, and
rises towards it as the tilt angle decreases. Furthermore, our
calculations indicate that the GB formation energy behaves
non-monotonically with the tilt angle, due to a change in the the
spatial distribution and relative contributions of the bond-stretching
and bond-bending deformations associated with the formation of the
defect.

\begin{acknowledgments}
We acknowledge support from the Brazilian agencies CNPq, FAPEMIG, Rede
de Pesquisa em Nanotubos de Carbono, INCT de Nanomateriais de Carbono,
and Instituto do Mil\^enio em Nanotecnologia-MCT.
\end{acknowledgments}


\begin{thebibliography}{100}

\bibitem{rmp} A.~H.~C.~Neto {\textit et al.}, Rev.~Mod.~Phys.
{\bf 81}, 109 (2009), and references therein.

\bibitem {novos07} A.~K.~Geim and K.~S.~Novoselov, Nat.~Mater. {\bf
06}, 183 (2007).

\bibitem{nakada} K.~Nakada, M.~Fujita, G.~Dresselhaus, and
M.~S.~Dresselhaus, Phys.~Rev.~B {\bf 54}, 17954 (1996).

\bibitem{wallace} P.~R.~Wallace, Phys.~Rev {\bf 71}, 622 (1947).

\bibitem{novos04} K.~S.~Novoselov {\textit et al.}, Science {\bf 306},
 666 (2004).

\bibitem{kats06} M.~I.~Katsnelson, K.~S.~Novoselov, and A.~K.~Geim,
Nat.~Phys. {\bf 2}, 620 (2006).

\bibitem{deheer} C.~Berger {\textit et al.}, J. Phys. Chem. B {\bf
108}, 19912 (2004); J.~Hass {\textit et al.}, Phys.~Rev.~B 
{\bf 75}, 214109 (2007).

\bibitem{sinitsyna} O.~V.~Sinitsyna and I.~V.~Yaminsk,
Russ.~Chem.~Rev. {\bf 75}, 23 (2006).

\bibitem{laura} L.~B.~Biedermann, M.~L.~Bolen, M.~A.~Capano,
D.~Zemlyanov, and R.~G.~Reifenberger, Phys.~Rev.~B {\bf 79}, 125411
(2009).

\bibitem{datta} S.~S.~Datta, D.~R.~Strachan, E.~J.~Mele, and
A.~T.~C.~Johnson, Nanoletters {\bf 9}, 7 (2009).

\bibitem{simonis} P.~Simonis {\textit et al.}, Surf.~Sci. {\bf 511}, 319 (2002).
\bibitem{gan} Y.~Gan, W.~Chu, and L.~Qiao, Surf.~Sci. {\bf 539}, 120 (2003).

\bibitem{pong} W.-T.~Pong, J.~Bendall, and C.~Durkan, Surf.~Sci. {\bf
601}, 498 (2007).

\bibitem{varchon} F.~Varchon, P.~Mallet, L.~Magaud, and
J.-Y.~Veuillen, Phys.~Rev.~B {\bf 77}, 165415 (2008).

\bibitem{cervenka} J.~Cervenka and C.~F.~J.~Flipse, Phys.~Rev.~B
{\bf 79}, 195429 (2009). 

\bibitem{gong} G.~Gu {\textit et al.}, Appl.~Phys.~Lett. {\bf 90},
253507 (2007).

\bibitem{Kohn} W.~Kohn and L.~J.~Sham, Phys.~Rev. {\bf 140}, A1133 (1965).

\bibitem{Kleinman} L.~Kleinman and D.~M.~Bylander,
Phys.~Rev.~Lett. {\bf 48}, 1425 (1982).

\bibitem{Troullier} N.~Troullier and J~.L.~Martins, Phys.~Rev.~B 
{\bf 43}, 1993 (1991).

\bibitem{Siesta} J. M. Soler {\textit et al.}, J.~Phys.~Cond.~Matt. {\bf 14}
2745 (2002).

\bibitem{park} C.~H.~Park, L.~Yang, Y.~W.~Son, M.~L.~Cohen,
S.~G.~Louie, Phys.~Rev.~Lett.  {\bf 101}, 126804 (2008).

\bibitem{vozmed} A.~Carpio, L.~L.~Bonilla, F.~de~Juan,
M.~A.~H.~Vozmediano, New J. of Phys. {\bf 10}, 053021 (2008).

\end{thebibliography}
\end{document}